\documentclass[conference]{IEEEtran}
\IEEEoverridecommandlockouts
\usepackage{cite}
\usepackage{amsmath,amssymb,amsfonts, physics, dsfont}
\usepackage{algorithmic}
\usepackage{graphicx}
\usepackage{textcomp}
\usepackage{xcolor}
\def\BibTeX{{\rm B\kern-.05em{\sc i\kern-.025em b}\kern-.08em
    T\kern-.1667em\lower.7ex\hbox{E}\kern-.125emX}}
\begin{document}

\title{Coupling Quantum Antennas to Fibers and Waveguides}

\author{\IEEEauthorblockN{Girish S. Agarwal\textsuperscript{1,2}}
\IEEEauthorblockA{\textit{\textsuperscript{1}Institute for Quantum Science and Engineering,} \\}
\IEEEauthorblockA{\textit{\textsuperscript{2}Department of Biological and Agricultural Engineering,} \\
\textit{Texas A\&M University,}\\
College Station, USA.\\
girish.agarwal@tamu.edu}
\and
\IEEEauthorblockN{Debsuvra Mukhopadhyay}
\IEEEauthorblockA{\textit{Department of Physics and Astronomy,} \\
\textit{Texas A\&M University,}\\
College Station, USA.\\
debsosu16@tamu.edu}
}

\maketitle

\begin{abstract}
We present a brief overview of the transport of quantum light across a one-dimensional waveguide which is integrated with a periodic string of quantum-scale dipoles. We demonstrate a scheme to implement transparency by suitably tuning the atomic frequencies without applying a coupling field and bring out the pronounced non-reciprocity of this optical device. The fiber-mediated interaction between integrated dipoles allows one to achieve both dispersive and dissipative couplings, level repulsion and attraction, and enhanced sensing capabilities. All these ideas can be translated to a wide variety of experimental setups of topical interest such as resonators on a transmission line, cold atoms near a fiber and quantum dots coupled to plasmonic excitations in a nanowire or photonic crystal waveguides\footnote{Manuscript is based on an invited contribution to the special session on ``Quantum Antennas and Photonic Quantum Sensing" at IEEE COMCAS 2021, Israel.}.
\end{abstract}


\section{Introduction}
\begin{figure*}
\begin{center}
\includegraphics[width=170mm, height=60mm]{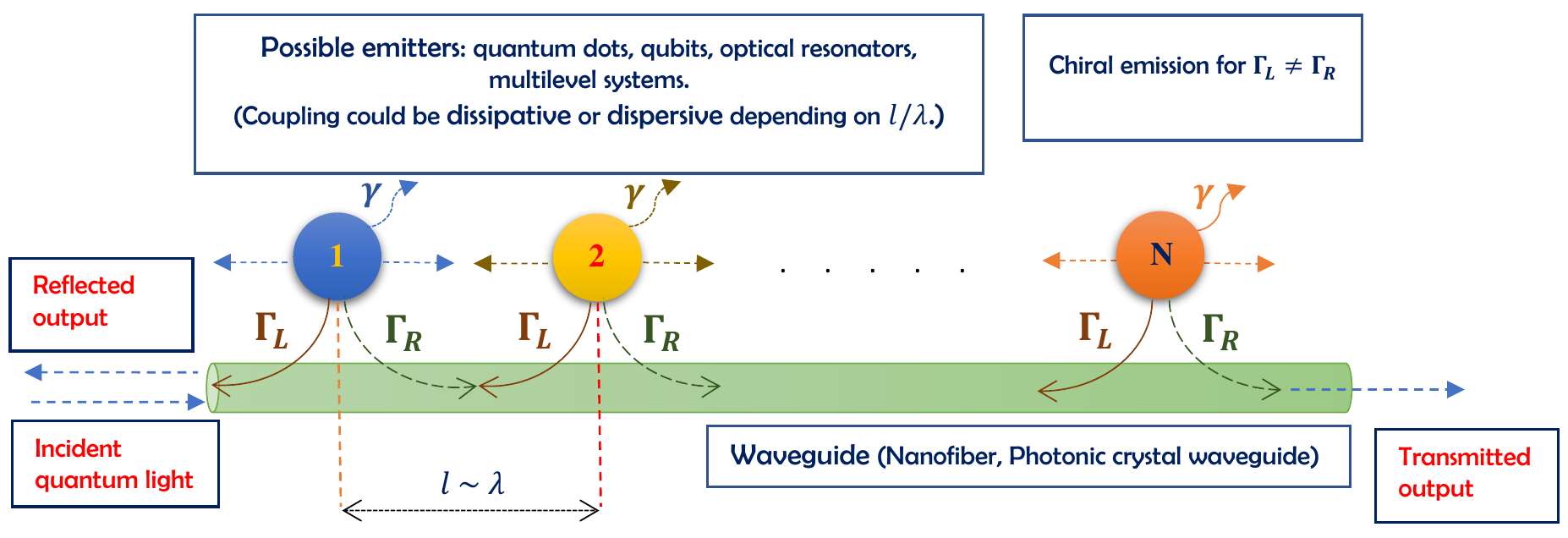}
\end{center}
\caption{Schematic of a multiemitter lattice evanescently coupled to a one-dimensional (1D) waveguide. The interemitter separation $l$ is scale-wise similar to or larger than the wavelength of the guided photon.}
\label{f1}
\end{figure*}
Quantum antennas (QA) can be considered as atomic scale dipoles coupled to radiation fields in different environments, such as plasmonic structures, nanofibers, and waveguides. A modification of spontaneous emission \cite{b1,b2} from atomic level dipoles, which could be atoms, quantum dots and other qubits and even higher dimensional systems such as qutrits and qudits, becomes a probe of the electromagnetic fields. The strong coupling of QA to such structures, being fundamentally important, yields new ways of probing electromagnetic fields and the physical structures producing such fields. The QAs can also get strongly coupled to each other with coupling mediated by the electromagnetic environment \cite{b3} leading to quantum entanglement \cite{b4}. The QAs in anisotropic environment can exhibit a variety of interference effects \cite{b2,b5}.

\section{Quantum Antennas Coupled To Fibers and Waveguides}
Waveguide QED offers a complementary paradigm to cavity QED for studying and regulating light-matter interaction which forms the cornerstone of state-of-the-art chip-scale photonics \cite{bnew}. Nanofiber-controlled fluorescence from an evanescently coupled atom was first demonstrated by Le Kien \textit{et al} \cite{b6}. The radiative behavior of multi-emitter configurations is governed by the relative location of the emitters, allowing more efficient control on photon transport and possible adoption as a prototypical element for quantum information. An ensemble of emitters interfacing with a waveguide, as shown in Fig. \ref{f1}, exhibits a myriad of exotic phenomena such as single-photon super- and sub-radiance, non-reciprocal photon transport, and asymmetrical Fano lineshapes \cite{b7}. By stimulating interaction between photons, a multi-spin cluster can serve as an optical switch or nearly any possible quantum gate. Possibility of transparency due to single-photon transport across differentially detuned two-level emitters has been theoretically demonstrated, without applying any control field \cite{b8}. Directional dependence of the emission can entail asymmetric relaxation rates to the left- and right-propagating modes, which significantly modifies the transport. The interaction mediated by an optical fiber wields flexible control on both dispersive and dissipative couplings, level repulsion and attraction, and on sensing capabilities \cite{b9}.  These ideas can be contextualized in a variety of experimental models of topical interest such as resonators attached to a transmission channel, cold atoms trapped near a fiber and quantum dots coupled to a nanowire or photonic crystal waveguides \cite{b10}.  Fiber-mediated coherence between emitters is instrumental to the sensing potential of these devices and can also generate long-range interemitter entanglement. In order to deftly explore topological phenomena in qubits coupled to fiber waveguides, more technical sophistication based on modern nanofabrication methods is needed to increase the qubit-waveguide coupling. In what follows, we bring out some intriguing results on photonic transport through a waveguide-integrated dipolar lattice, which are based partly on a previous publication \cite{b8}.

\section{Multiemitter Transparency Effects}
When a periodic chain of two-level emitters is strongly coupled to a waveguide, the spatial periodicity considerably impacts the spectral behavior, which can be attributed to a waveguide-mediated phase coupling between the atoms \cite{b7}. Such interaction effects exist even when the interatomic separations are large enough to effectively nullify dipole-dipole couplings. Here, we analyze single-photon transport through a 1D waveguide. Our system consists of a periodic array of $N$ two-level emitters evanescently coupled to a 1D continuum (like in Fig. \ref{f1}), with the coupling strength signified by $G$. For a single-photon source, our analytical model can be restricted to the single-photon manifold of the relevant Hilbert space. We assume the interatomic separation $l$ to be comparable to or larger than the resonance wavelength, enabling us to discard dipole-type interactions between the atoms. One can solve for the scattering eigenstate using the Schr$\ddot{\text{o}}$dinger equation and obtain the spectral amplitudes (reflection and transmission) via the transfer matrix method as
\begin{align}
r=\frac{(\prod_{j=1}^NL_j)_{12}}{(\prod_{j=1}^NL_j)_{22}},\hspace{2mm}t=\frac{e^{-iNkl}}{(\prod_{j=1}^NL_j)_{22}},
\end{align}
where $L_j=\begin{bmatrix}\\
e^{ikl}(1-i\delta_j^{-1})&-ie^{-ikl}\delta_j^{-1}\\
ie^{ikl}\delta_j^{-1}&e^{-ikl}(1+i\delta_j^{-1})\\
\end{bmatrix}$
is the transfer matrix  pertaining to the reflection/transmission across the $j^{th}$atom, $k$ is the wavelength of the incident photon, $\delta_j=(v_g k-\omega_j+i\Gamma_0)/\Gamma$ is a scaled detuning of the $j^{th}$ atom, $\Gamma=G^2/v_g$ , $\Gamma_0$ describes dissipation outside of the waveguide channel, and $v_g$ is the group velocity of transport. The matrix product becomes tractable when the lattice periodicity becomes commensurate with the wavelength, i.e., for $kl = n\pi$, with $n$ as any natural number. In this case, the transfer matrices reduce to $L_j=(-1)^n [\mathds{1}+\delta_j^{-1} \rho_-]$, where $\mathds{1}$ is the $2$x$2$ Identity matrix, $\rho_-=\sigma_y-i\sigma_z$ is a spin-annihilation operator in the $\sigma_x$-eigenbasis. Since $\rho_-\rho_-=0$, all transfer matrices commute, and the transfer matrix product becomes $\prod_{j=1)}^NL_j=(-1)^nN [\mathds{1}+\sum_{j=1}^N\delta_j^{-1}\rho_-]$. The spectral coefficients are obtained to be
\begin{align}
r&=-\frac{i\Gamma\sum_{j=1}^N(\Delta_j+i\Gamma_0)^{-1}}{1+i\Gamma\sum_{j=1}^N(\Delta_j+i\Gamma_0)^{-1}},\notag\\
t&=\frac{1}{1+i\Gamma\sum_{j=1}^N(\Delta_j+i\Gamma_0)^{-1}}.
\end{align}
\begin{figure*}
\begin{center}
\includegraphics[width=170mm, height=95mm]{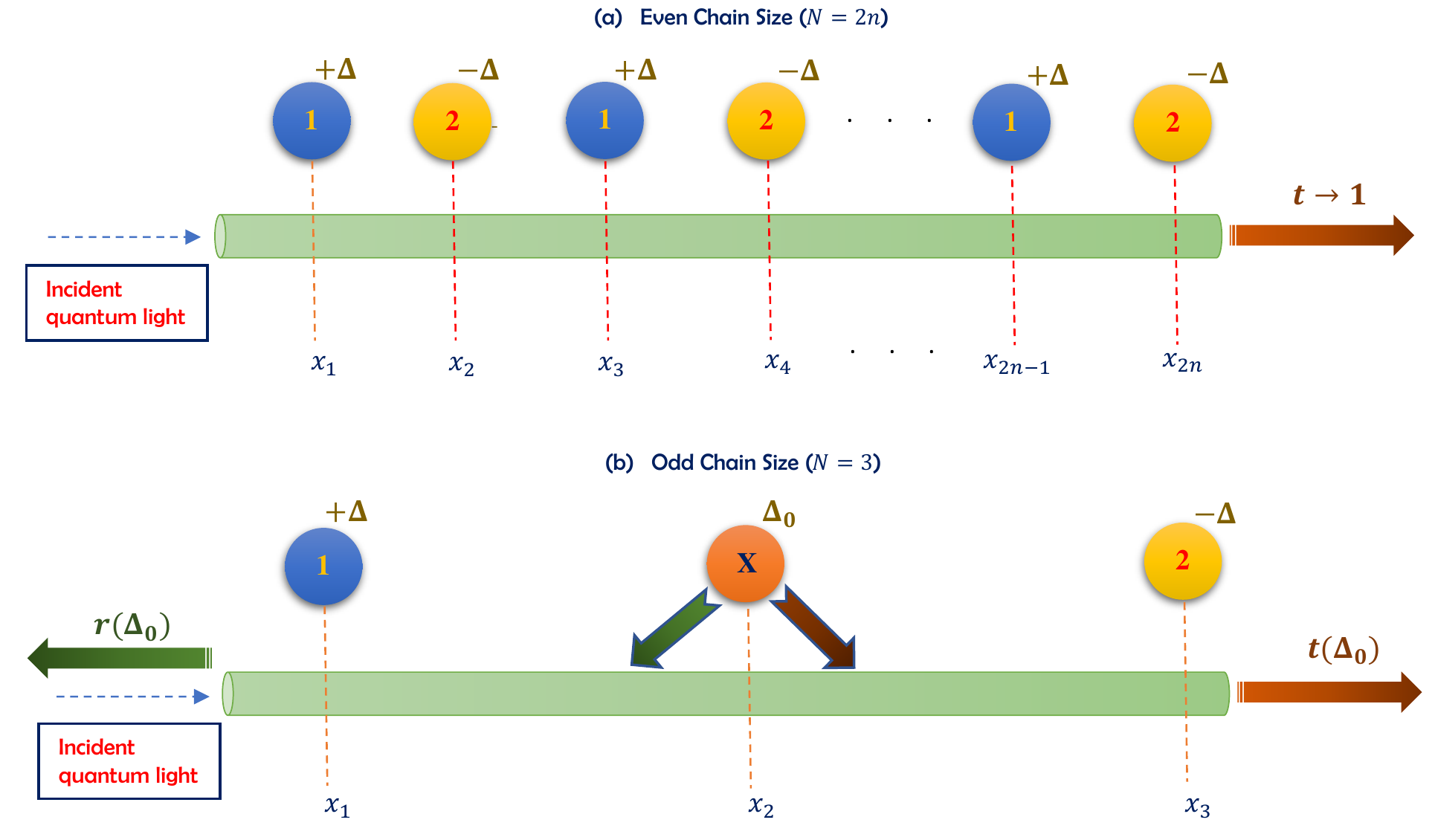}
\end{center}
\caption{(a) Even number of emitters with equal and opposite detunings assigned in pairs generates transparency. The order shown here is one of the simplest permutations. (b) A chain of three atoms, out of which two carry equal and opposite detunings $+\Delta$ and $-\Delta$, and the odd one out, marked by $X$ carries a detuning $\Delta_0$. The latter solely determines the spectral behavior. This character holds generally for any odd chain size with a similar assignment of detunings.}
\label{f2}
\end{figure*}
If all the atoms bear identical frequencies, the cooperative emission retains its Lorentzian profile with a decay rate scaling linearly to the chain size, resulting in Dicke-type superradiance. Under more specific conditions when they are all resonant with the incident photon, we get $t=\Gamma_0/(\Gamma_0+N\Gamma)$, signifying near-perfect reflection in the limit $\Gamma_0\ll\Gamma$. For non-identically detuned atoms, the parity of the chain size, i.e., whether the chain size is even or odd, is found to play a crucial role on the control of collective radiance \cite{b8}. For an even parity, if the atoms are detuned pairwise oppositely in signature but equally in magnitude, $t$ approaches $1$ when $\Gamma_0\ll\Gamma$. Rigorously stated, if there are $N = 2n$ atoms, an assignment of the frequency detunings $+\Delta^{(1)}, -\Delta^{(1)}, +\Delta^{(2)}, -\Delta^{(2)}, ... +\Delta^{(n)}, -\Delta^{(n)}$, in no particular order, would give rise to transparency in the system. On the other hand, for an odd chain size with sufficiently small dissipation, one can recover the single-atom emission spectra by assigning pairwise asymmetric detunings to any randomly chosen $(N-1)/2$ emitter pairs, leaving out a single atom which entirely governs the spectral characteristics. The two schemes are illustrated in Fig. \ref{f2}.\\

\textit{Transparency with emitters in a cavity:} One could also consider a chain of $N$ atomic scatterers placed at the antinodes of an optical cavity. An optical laser of frequency $\omega_{\text{laser}}$ drives the cavity, which, then, excites the emitters coupled to the intracavity field. Assuming a cavity leakage rate of $\kappa$ across each mirror, while ignoring internal cavity losses and atomic dampings, the transmission across the multiemitter-cavity system turns out to be
\begin{align}
t=\frac{2\kappa}{2\kappa+i\bigg(\delta-g^2\sum_{j=1}^N\frac{1}{\Delta_j}\bigg)},
\end{align}
where $\delta=\omega_{\text{cavity}}-\omega_{\text{laser}}$, $g$ quantifies the strength of atom-cavity coupling, $\Delta_j=\omega_j-\omega_{\text{laser}}$ is the detuning of the $j^{th}$ atom. For simplicity, if we set $\delta=0$, we recover the results akin to the waveguide transport. When the atoms have identical detunings, superradiant effect becomes evident from Eq. (3), as the collective linewidth becomes $Ng^2/(2\kappa)$. For disparate detunings, the possibility of transparency hinges on the parity of the ensemble size. An even chain size allows transparency upon imparting equal and opposite pairwise detunings, while an odd chain size, for a similar assignment protocol, leads to an effective retrieval of single-atom emission characteristics. Thus, there are obvious parallels between cavities and waveguides when they are integrated with quantum antennas.

\section{Non-reciprocity of Photon Transport}
\begin{figure*}
\begin{center}
\includegraphics[width=165mm, height=70mm]{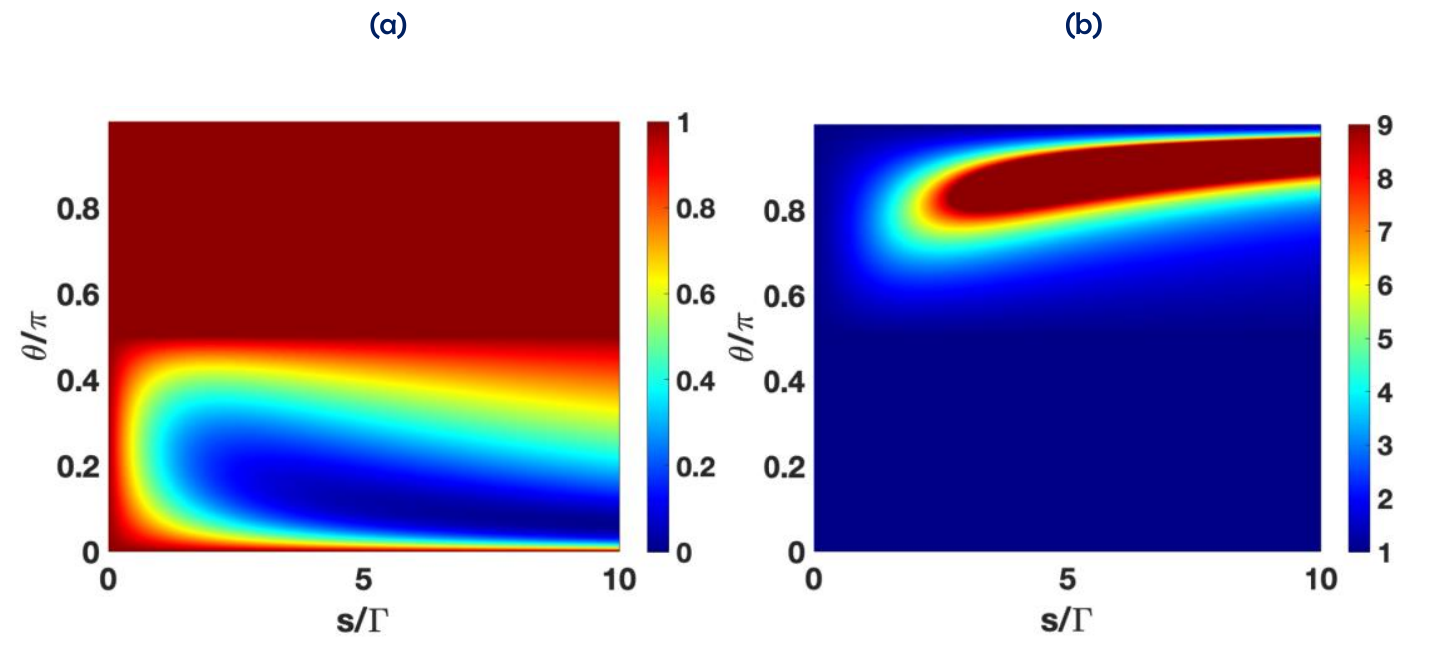}
\end{center}
\caption{Degree of non-reciprocity, $\eta$, as a bivariate function of $\theta=kl$ and $s/\Gamma$. The left graph (a) prominently captures the region where $\eta<1$, while the right one (b) accentuates the spot in which $\eta<1$. All values in the range $\eta\geqslant 1$ are encoded as $1$ in (a), while those in the set $\eta\geqslant 9$ are mapped to $9$ in (b). Very small or very large values of $\eta$ indicate substantial non-reciprocity. System parameters have been constrained to the space $\overline{\Delta}=0$ (i.e., equal and opposite detunings), and $\Gamma=\Gamma_0$ (i.e. critical coupling).}
\label{f3}
\end{figure*}

 For general system parameters, the noncommutativity of transfer matrices precludes the possibility of having reciprocal photon transport through the waveguide, even while there is symmetric coupling between the atoms and the waveguide. However, when $kl$ is a multiple of $\pi$, one has perfect reciprocity \cite{b8}. Let us consider the simpler scenario of two differentially detuned atoms in a waveguide and define the mean detuning of the incident photon $\overline{\Delta}=v_g k-(\omega_1+\omega_2)/2$ and the relative atomic detuning $s=\omega_1-\omega_2$. Then, for the general phase-dependent paradigm, the reflection from the atomic system can be expressed as
\begin{align}
r=-\frac{\Gamma^2(e^{i\alpha}-1)+i\Gamma[(e^{i\alpha}+1)(\overline{\Delta}+i\Gamma_0)-(e^{i\alpha}-1)s/2]}{(\overline{\Delta}+i(\Gamma+\Gamma_0))^2+\Gamma^2e^{i\alpha}-(s/2)^2},
\end{align}
where $\alpha=2kl$. Therefore, unless $kl$ is an integral multiple of $\pi$, the above expression depends on $s$ which is order dependent. To quantify the extent of non-reciprocity, we could define the ratio $\eta=\abs{\frac{r_{12}}{r_{21}}}^2$, where $r_{ij}$ represents the coefficient of reflection from an atomic chain with the atoms with frequency $\omega_i$ and $\omega_j$ in positions $1$ and 2 respectively. Clearly, $\eta=1$ would signify perfect reciprocity. From Eq. (4), it follows, on using $(e^{i\alpha}-1)/(e^{i\alpha}+1)=i\tan(\alpha/2)$, that
\begin{align}
\eta=\frac{(\overline{\Delta}+\Gamma\tan kl)^2+(\frac{s}{2}\tan kl-\Gamma_0)^2}{(\overline{\Delta}+\Gamma\tan kl)^2+(\frac{s}{2}\tan kl+\Gamma_0)^2},
\end{align}
which deviates from unity whenever $\Gamma_0\neq0$ and $kl = n\pi$. Thus, the existence of a dissipative channel is central to the absolute reciprocity of photonic energy transfer. Note that Eq. (4) also implies $\eta(-s)=[\eta(s)]^{-1}$, which means that if a certain $s$ maximizes $\eta$, flipping the sign of $s$ would minimize the same for an otherwise constant set of parameters. The two panels in Fig. \ref{f3} demonstrate the two-dimensional plots of $\eta$ when $\overline{\Delta}=0$ and $\Gamma=\Gamma_0$, with the convention $s>0$. The value of $\eta$ can be much larger than 1 as well if it lies well within the dark red regions shown on either panel. For instance, the phase $kl=5\pi/6$ supports a maximum asymmetry of $\eta\approx13.78$ in the chosen parameter space. This indicates that the reflected intensity for one direction of incidence is almost $14$ times that for the opposite direction of incidence. 

\section{Conclusions}
With the volume of progress being made in waveguide QED, waveguide-integrated circuits and networks are gaining more attention in the context of purely optics-based quantum communication. In this article, we have analyzed single-photon transmission through a 1D waveguide strongly coupled to a periodic array of quantum antennas modeled as harmonic or two-level emitters. The multitude of emission pathways in the photonic interaction with an ensemble of emitters makes the photon transport strongly non-reciprocal and sensitive to the spatial separation. By tuning the frequencies of emitters interfacing with a common fiber, we exhibit a feasible protocol for achieving transparency without using any control field. If the atoms are differentially detuned, non-reciprocity ensues as a combined effect of the spatial separation and dissipation into extraneous channels. The asymmetry of photon transport exists even in the absence of chiral couplings whereby either of the emitters couples differently to the two oppositely traveling light modes in the waveguide continuum. The non-reciprocity can be hugely beneficial to the design of optical diodes which permit signal transport in one direction while stemming the flow in the other. We also demonstrated transparency effects in multiatom-integrated cavity systems and brought out analogies with the transport of single photons through waveguides.

\section*{Acknowledgment}
The authors acknowledge the support of The Air Force Office of Scientific Research [AFOSR award no FA9550-20-1-0366], The Robert A. Welch Foundation [grant no A-1943] and the Herman F. Heep and Minnie Belle Heep Texas A\&M University endowed fund.

\end{document}